\documentclass[letterpaper]{jpconf}
\usepackage{graphicx}
\begin{document}

\title{Prospect for measuring the branching ratio of $B_{s}\rightarrow\mu\mu$ at LHC{\it b}}

\author{EL\'IAS L\'OPEZ ASAMAR}
\address{Universitat de Barcelona\\
  Avinguda Diagonal 647, 08028 Barcelona (Spain)\\
  E-mail: elopez$@$ecm.ub.es
}

\author{ON BEHALF OF THE LHC{\it b} COLLABORATION}

\begin{abstract}
The Standard Model predicts a branching ratio for the decay mode $B_{s}\rightarrow\mu\mu$ of (3.32$\pm$0.32)$\times$10 $^{-9}$ while some SUSY models predict enhancements of up to 2 orders of magnitude. It is expected that at the end of its life the Tevatron will set an exclusion limit for this branching ratio of the order of 10 $^{-8}$, leaving one order of magnitude to explore. The efficient trigger, excellent vertex reconstruction and invariant mass resolution, and muon identification of the LHC{\it b} detector makes it well suited to observe a branching ratio in this range in the first years of running of the LHC.
In this article an overview of the analysis that has been developed for the measurement of this branching ratio is presented. The event selection and the statistical tools used for the extraction of the branching ratio are discussed. Special emphasis is placed on the use of control channels for calibration and normalization in order to make the analysis as independent of simulation as possible. Finally, the expected performance in terms of exclusion and observation significance are given for a set of values of integrated luminosities.
\end{abstract}

\section{Introduction}
In the Standard Model (SM), flavour changing neutral currents (FCNC) can be generated only through loop diagrams, resulting in low branching ratio (BR) predictions. Some models beyond SM can introduce contributions of size comparable to that of SM, yielding significantly different predictions for these branching ratios.
\newline 
\indent
This is the case of the decay $B_{s}\rightarrow\mu\mu$, suppressed by helicity, for which the SM predicts BR($B_{s}\rightarrow\mu\mu$) = (3.32$\pm$0.32)$\times$10 $^{-9}$ \cite{SM-prediction}. In the Minimal Supersymmetric extension of SM (MSSM) this quantity is proportional to tan$^6\beta$ \cite{MSSM-prediction}, leading to enhancements of up to one order of magnitude in cases such as the Non-Universal Higgs Masses framework (NUHM) \cite{NUHM-prediction}.
\newline
\indent
The current experimental limit for this BR is BR($B_{s}\rightarrow\mu\mu$) $<$ 47$\times$10$^{-9}$ at 90\% C.L. \cite{CDF-measurement} (CDF collaboration, with 2 fb$^{-1}$), roughly a factor 15 above the SM prediction. This channel then offer the possibility of observing hints of Physics beyond SM, or in case of confirming the SM prediction, rejecting an important region of the parameter space of some of these models.

\section{Experimental conditions}
The LHC{\it b} detector \cite{TDR} is specially designed for studying $B$ meson decays produced in LHC from $pp$ collisions at a centre of mass energy of $\sqrt{s}$ = 14 TeV/c$^2$). The nominal integrated luminosity at the LHC{\it b} interaction point will be 2 fb$^{-1}$/year, resulting in $\sim$4$\times$10$^{11}$ $B$ meson pairs per year inside the acceptance of the detector. The performances most relevant for the measurement are the vertexing capabilities to identify the displaced vertices of $B$ decays (impact parameter (IP) resolution of 14+35/p$_T$ $\mu$m, p$_T$ standing for the transverse component of the momentum), the invariant mass resolution \cite{FlavLHC} ($\sim$20 MeV/c$^2$, compared to $\sim$35 MeV/c$^2$ at CMS, or $\sim$80 MeV/c$^2$ at ATLAS), and the muon identification \cite{muonID} (94\% efficient with 1\% pollution coming from mis-identified $\pi$/$K$).

\section{Analysis strategy}
\begin{figure}
\begin{center}
\includegraphics[width=0.5\textwidth]{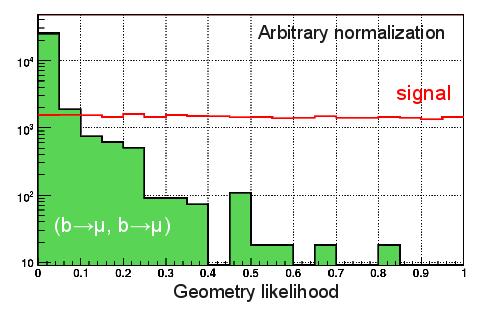}
\caption{Distribution of geometrical likelihood for the signal (red line) and background coming from $B$ decays (green, filled), satisfying only the selection cuts (no trigger is applied).}
\label{GL}
\end{center}
\end{figure}

The first step of the analysis consist of a very loose selection that focus on efficiency for signal events, rather than on background rejection. This preliminary selection is based on the quality of the vertex, the invariant mass, the impact parameter (IP) of the muons, and the IP/$\sigma_{IP}$, decay length and momentum of the reconstructed $B$. It is 65\% efficient over reconstructed signal events, keeping the bulk of the sensitive signal, and yields $\sim$40 signal events (assuming SM predictions) and $\sim$13$\times$10$^4$ background events per fb$^{-1}$ under nominal conditions, inside a mass window of 60 MeV/c$^2$ around the mass of the $B_s$.
\newline
\indent
The remaining events are caracterized according to three discriminant variables \cite{bsmumu}: 

\begin{itemize}
\item Invariant mass. 
\item Particle-ID likelihood (PIDL) which for both muons combines the likelihood of being muon and the likelihood of not being muon.
\item Geometrical likelihood (GL) which combines the information of the distance of closest approach of the tracks of the two muons, the minimal IP/$\sigma_{IP}$ of the muons, the isolation of both muons, and the lifetime and impact parameter of the $B$ \cite{bsmumu}. Its distribution for signal and background is shown in Figure \ref{GL}.
\end{itemize}

\indent
Backgrounds are distributed differently from signal in the three-dimensional space spanned by the invariant mass and both likelihoods, thus defining regions of different sensitivity to the signal. Only backgrounds lying in sensitive regions can affect significantly the measurement. Defining the sensitive events as those having GL $>$ 0.5, the background composition of this subsample is shown in Table \ref{backgrounds} \cite{bsmumu}. In this case, the signal-to-background ratio rises from 0.3\% to 11\%.

\begin{table}
\begin{center}
\begin{tabular}{|c|c|}
\hline
Channel & Yield (2 fb$^{-1}$) \\
\hline
\hline
$(b\rightarrow\mu,b\rightarrow\mu)$ & 170 $\pm$ 90 \\
\hline
$B_{c}^{+}\rightarrow J/\Psi (\mu\mu)\mu^{+}\nu_{\mu }$ & $<$ 20 (90\% C.L.) \\
\hline 
$B\rightarrow h^+h^-$ mis-ID & 8 $\pm$ 2 \\
\hline
\hline
Signal & 20 $\pm$ 2 \\
\hline
\end{tabular}
\caption{Sources of background for sensitive events (GL $>$ 0.5).}
\label{backgrounds}
\end{center}
\end{table}

\section{Calibration and normalization}
Calibration and normalization procedures foreseen for this analysis rely entirely on data, in order to minimize dependence on simulations \cite{bsmumu}. Mass sidebands are used to calibrate background distributions, while control channels are used for calibration of signal properties:

\begin{itemize}
\item The particle-ID likelihood is calibrated using inclusive samples of $J$/$\Psi$($\mu\mu$) (for muon hypothesis) and $\Lambda$($pK$) (for non-muon hypothesis), which can be selected with high purity using only kinematical cuts.
\item The geometrical likelihood and invariant mass are calibrated using $B\rightarrow h^+h^-$, which has the same kinematical properties as the signal. The trigger introduces strong biases in the distribution of the geometrical likelihood, which can be removed by using events triggered on particles not related with the signal ($\sim$7\% of total $B\rightarrow h^+h^-$).
\end{itemize}

Control channels with known branching ratios allow to normalize the signal trough

\begin{equation}
N_S={\epsilon_S\over\epsilon_C}{f _{B,S}BR_S\over f _{B,C}BR_C}N_C \qquad .
\label{normalization}
\end{equation}

\noindent
Subindices $S$ and $C$ stand for signal and control channel respectively. The total efficiency $\epsilon$ can be split as $\epsilon_{rec/prod}\times\epsilon_{sel/rec}\times\epsilon_{trig/sel}$, where $\epsilon_{rec/prod}$ is the reconstruction efficiency on produced events, $\epsilon_{sel/rec}$ is the efficiency of the offline selection on reconstructed events, and $\epsilon_{trig/sel}$ is the efficiency of the trigger on offline-selected events. $B^0$ decays are aproppiate for calibration due to the accuracy in the measurements of their branching ratios. A proper choice of the control channel leads to cancellation of the effect of some sources of inefficiency:

\begin{itemize}
\item $B^+\rightarrow J/\Psi (\mu\mu )K^+$. The effect of the trigger is similar to that for the signal due to the $J$/$\Psi$ muons. The ratio of reconstruction efficiencies needs to take into account the different number of tracks in the final state between the signal and the control channel (two and three tracks respectively), and it is estimated through:
\begin{equation}
{ \epsilon _{rec} (2~tracks) \over \epsilon _{rec} (3~tracks) } \sim { \epsilon _{rec} (3~tracks) \over \epsilon _{rec} (4~tracks) }
\label{reco}
\end{equation}
Using $B^0\rightarrow J/\Psi (\mu \mu)K^{*0}(\pi K)$ as the four-track decay channel the ratio between the left-hand side and the right-hand side is 92\%.

\item $B\rightarrow h^+h^-$. Has the same kinematic properties to that of signal, leading to cancellation of reconstruction and selection effects. 
\end{itemize}

In both cases the ratio of trigger efficiencies is estimated using events triggered on particles not related with signal, which have a relatively small trigger bias on signal properties. 
\newline
\indent
The ratio of $B$ production fractions in Equation \ref{normalization} is the main source of uncertainty of the measurement, as long as channels coming from $B^0$ decays are used for normalization. It introduces a systematic error of 13\%.

\section{Extraction of the branching ratio}
\begin{figure}
\begin{center}
\includegraphics[width=\textwidth]{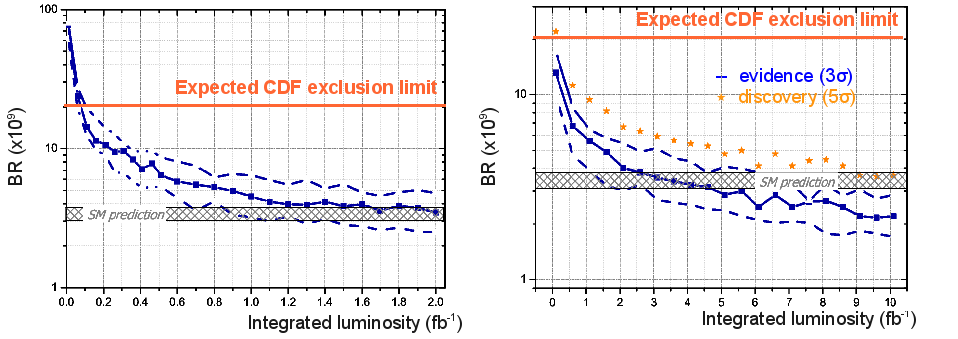}
\caption{Expected performance for exclusion (left) and measurement (right) of a given branching ratio as a function of the integrated luminosity. The expected limit set by CDF at the end of its life (8 fb$^{-1}$ of data) is also shown.}
\label{performance}
\end{center}
\end{figure}

The expected distributions of invariant mass and both particle-ID and geometrical likelihoods for signal and background will be obtained from the control channels and the mass sidebands respectively. These distributions will be combined together and compared with the measured ones, using the CL$_s$ method \cite{CLs}. In the absence of signal this analysis would reach the expected CDF limit with only an integrated luminosity of 0.1 fb$^{-1}$, and would reach the SM prediction with 2 fb$^{-1}$ \cite{bsmumu} (see Figure \ref{performance}). In case of observing a signal, the 3$\sigma$ (5$\sigma$) measurement could be achieved with 3 fb$^{-1}$ (10 fb$^{-1}$) \cite{bsmumu} of data.

\section{Conclusions}
The method developed for the measurement of the branching ratio of $B_{s}\rightarrow\mu\mu$ at LHC{\it b} makes an extensive use of the control channels in order to be independent of simulations. This analysis is expected to lead to relevant results even with early data, excluding the SM value at 90\% CL with only 2 fb$^{-1}$ in absence of signal, or observing it with only 3 fb$^{-1}$, thus potentially being able to provide one of the first evidences of New Physics from the LHC.
\newline

\end{document}